%% file: chalker.tex
\documentclass[vecphys]{svmult}

\usepackage{makeidx}         
\usepackage{graphicx}        
\usepackage{multicol}        
\usepackage{cite}            
\usepackage[bottom]{footmisc}

\makeindex             

\begin{document}

\title{ Geometrically frustrated antiferromagnets: statistical
  mechanics and dynamics}
\titlerunning{Fluctuations in geometrically frustrated magnets} 
\author{J. T. Chalker}
\institute{Theoretical Physics, Oxford University, 1, Keble Road,
  Oxford OX1 3NP, UK.
}

\maketitle

\section{Introduction}
\label{sec:1}
This chapter is intended to give an introduction to the theory 
of thermal fluctuations and their consequences for static and dynamic
correlations in geometrically frustrated 
antiferromagnets, focusing on the semiclassical limit,
and to discuss how our theoretical understanding leads to an
explanation of some of the main observed properties of these systems. 
A central theme will be the fact that simple, classical models for
highly frustrated magnets have a ground state degeneracy which is
{\it macroscopic}, though {\it accidental} rather than a consequence
of symmetries. We will be concerned in
particular with: (i) the origin of this degeneracy and the possibility
that it is lifted by thermal or quantum fluctuations; (ii) correlations
within ground states; and (iii) low-temperature dynamics. We
concentrate on Heisenberg models with large spin $S$, referring to the
chapter by G. Misguich for a discussion of quantum spin liquids, and
to the chapter by M. Gingras for an overview of
geometrically frustrated Ising models in the context of spin ice
materials. Several earlier reviews provide
useful further reading, including \cite{ramirez94},
\cite{schiffer96} and \cite{harris96} for experimental background, and
\cite{moessner01} and \cite{henley01} for an alternative
perspective on theory.

To provide a comparison, it is useful to begin by recalling
the behaviour of an {\it unfrustrated} antiferromagnet. To be
definite, consider the Heisenberg model with nearest neighbour
exchange $J$ on a simple cubic lattice. As the lattice is bipartite
-- it can be separated into two interpenetrating sublattices, in such
a way that sites of one sublattice have as their nearest neighbours
only sites from the other sublattice -- the classical ground states
are two-sublattice N\'eel states, in which spins on one sublattice all
have the same orientation, and those on the other sublattice have the opposite
orientation. These states are unique up to global spin rotations, which
are a symmetry of the model. Their only low energy excitations are long
wavelength spinwaves. These are Goldstone modes -- a consequence of
the symmetry breaking in ground states -- and have a frequency
$\omega({\bf k})$ that is linear in wavevector $\bf k$ at small
$k$. This classical picture carries over to the quantum system, and
for $S\gg 1$ it is sufficient to take account of 
fluctuations using harmonic spinwave theory. In particular, within
this approximation the sublattice magnetisation at low temperature is
reduced from its classical ground state value $S$ by an amount
\begin{equation}
\delta S = \frac{1}{\Omega} \int_{\rm BZ}
 \,\frac{zJS}{\hbar \omega({\bf k})}\left[\langle n({\bf k})\rangle
  +1/2\right]{\rm d}^3{\bf k} -\frac{1}{2}\;,
\label{sublattice}
\end{equation}
where $\langle n({\bf k})\rangle$ is a Bose factor giving the number
of thermally excited spin waves at wavevector $\bf k$ and $1/2$
represents the zero-point contribution, with the integral running
over the Brillouin zone of volume $\Omega$. Fluctuations increase with
temperature and the sublattice magnetisation falls to zero at the
N\'eel temperature $T_{\rm N}$. Within mean field theory ${\rm k}_{\rm
  B}T_{\rm N} = zJS^2$, where $z$ is the number of nearest neighbour
sites (six on the simple cubic lattice).

A central reason for the interest in geometrically frustrated magnets
is that they hold out the possibility of evading N\'eel
order. At the simplest level, there is a tendency for frustrated
systems to have many low-frequency modes,
which means both that excitations are
effective in reducing the ordered moment, because of the factor of
$zJS/\hbar \omega({\bf k})$ in (\ref{sublattice}), and that they are
thermally populated even at low temperature. More fundamentally, we
will see that frustration may lead to classical ground states of a
quite different kind and suppress $T_{\rm N}$ to zero.

\begin{figure}[t]
\centering
\includegraphics*[width=.4\textwidth]{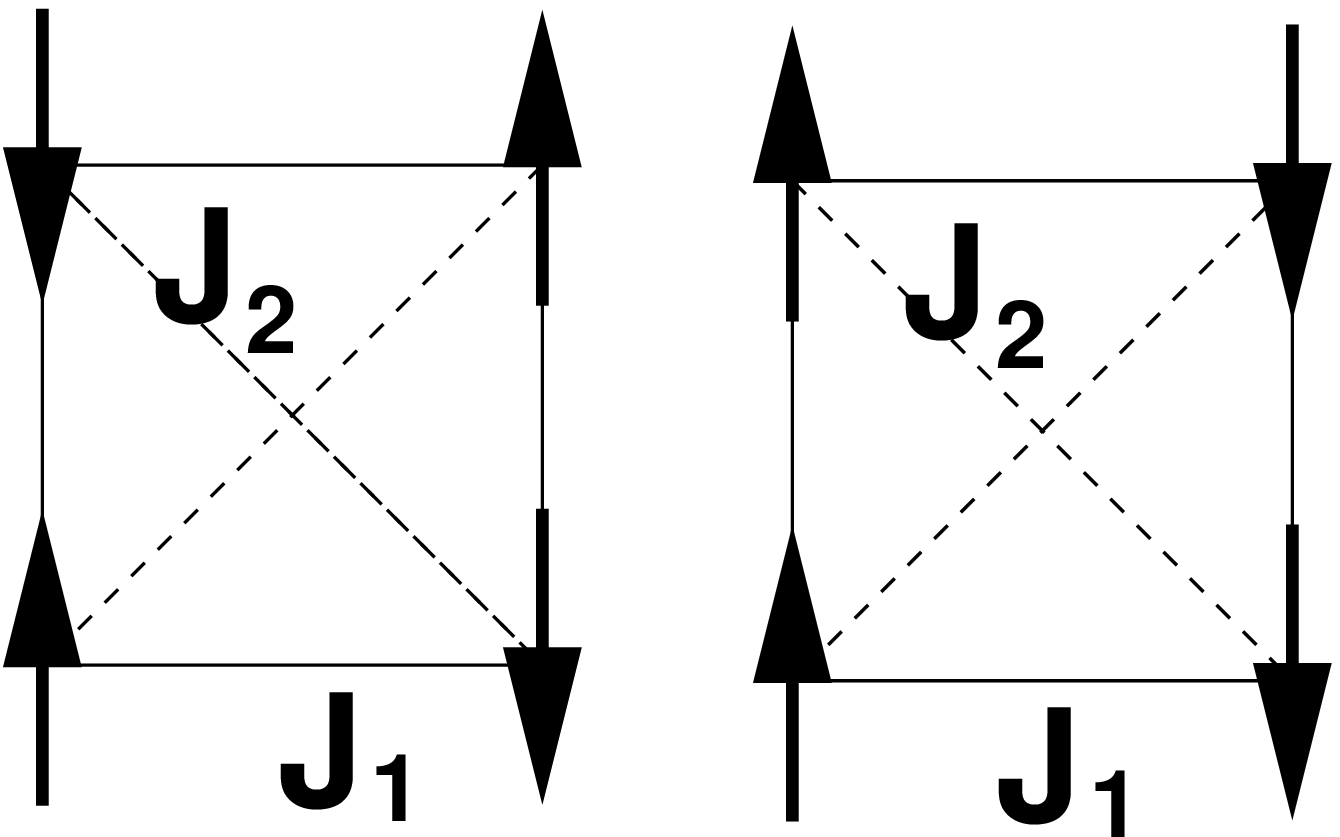} \qquad \qquad
\includegraphics*[width=.3\textwidth]{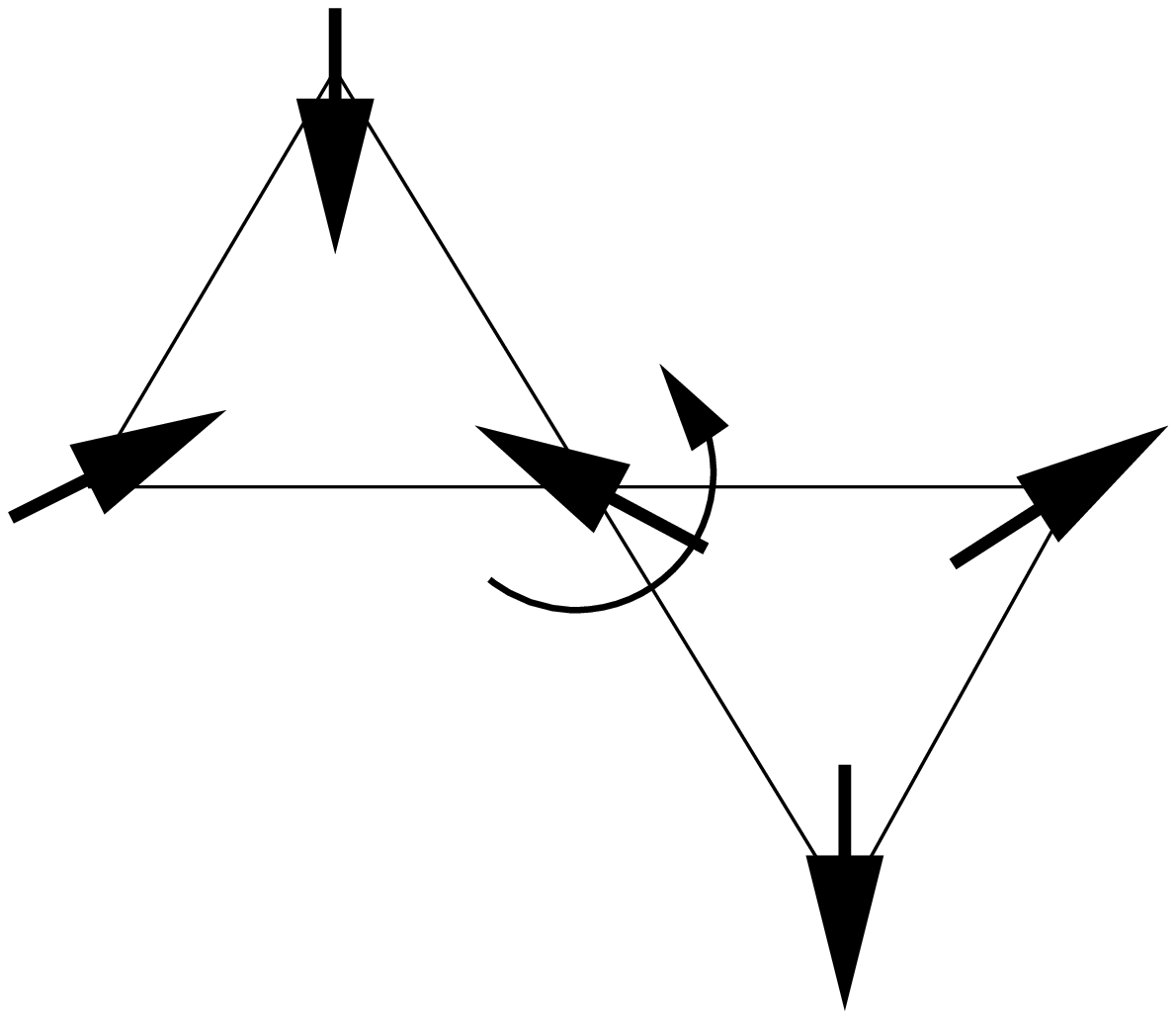}
\caption[]{Left and centre: $J_1-J_2$ model, showing ground state spin
configurations for: $2J_1>J_2$ (left); and $J_2> 2J_1$ (centre). Right:
ground states of classical Heisenberg spins at vertices of two
corner-sharing triangles, with degeneracy arising from rotations about
the common spin, as indicated.}
\label{fig1}       
\end{figure}

Since the term {\it frustration} is used in several different
contexts, it is worthwhile to set out some distinctions before going
further. In general terms, classical frustrated systems have
Hamiltonians with competing interactions which make contributions to
the energy that cannot simultaneously be minimised. The concept was
originally discussed in relation to spin glasses, but these are set apart from the
systems we are considering here by the fact that quenched disorder
has a controlling influence on their properties. Frustration as a way of destabilising
N\'eel order has been studied extensively in models with competing
nearest neighbour and further neighbour interactions, notably the
$J_1-J_2$ model on the square lattice \cite{chandra88}, illustrated in
Fig.~\ref{fig1}. The classical ground state of this model depends on
the ratio $J_1/J_2$. For $J_1 > 2J_2> 0$ neighbouring spins are
antiparallel, enforcing ferromagnetic alignment of second neighbours
and frustration of the interaction $J_2$. In the other regime, $2J_2
> J_1 >0$, second neighbours are antiferromagnetically aligned at the
expense of frustration of half of the $J_1$ interactions. Interest
focuses on the point $J_1=2J_2$ where these alternative classical
states are degenerate, and the consequences of frustration are likely
to be largest. While models of this kind provide an attractive
starting point for theoretical work, there are likely to be
difficulties in finding experimental realisations with interaction
strengths that place them close to the degeneracy point. From this
perspective, the long-appreciated \cite{anderson56,villain79}
attraction of geometrically frustrated magnets is that they are
systems in which structure alone may destabilise N\'eel order,
with only nearest neighbour interactions. To illustrate this at an
elementary level, consider Heisenberg spins at the vertices of two
corner-sharing triangles with nearest-neighbour antiferromagnetic
interactions, also shown in Fig.~\ref{fig1}. The ground states are
configurations in which spins within each triangle are coplanar and at relative angles of
$2\pi/3$. They have an accidental degeneracy (in
addition to that arising from symmetry) under relative rotations of
the spin planes for the two triangles about the axis defined by the
orientation of their common spin. 

\section{Models}
\label{sec:2}

\begin{figure}[h]
\centering
\includegraphics*[width=.3\textwidth]{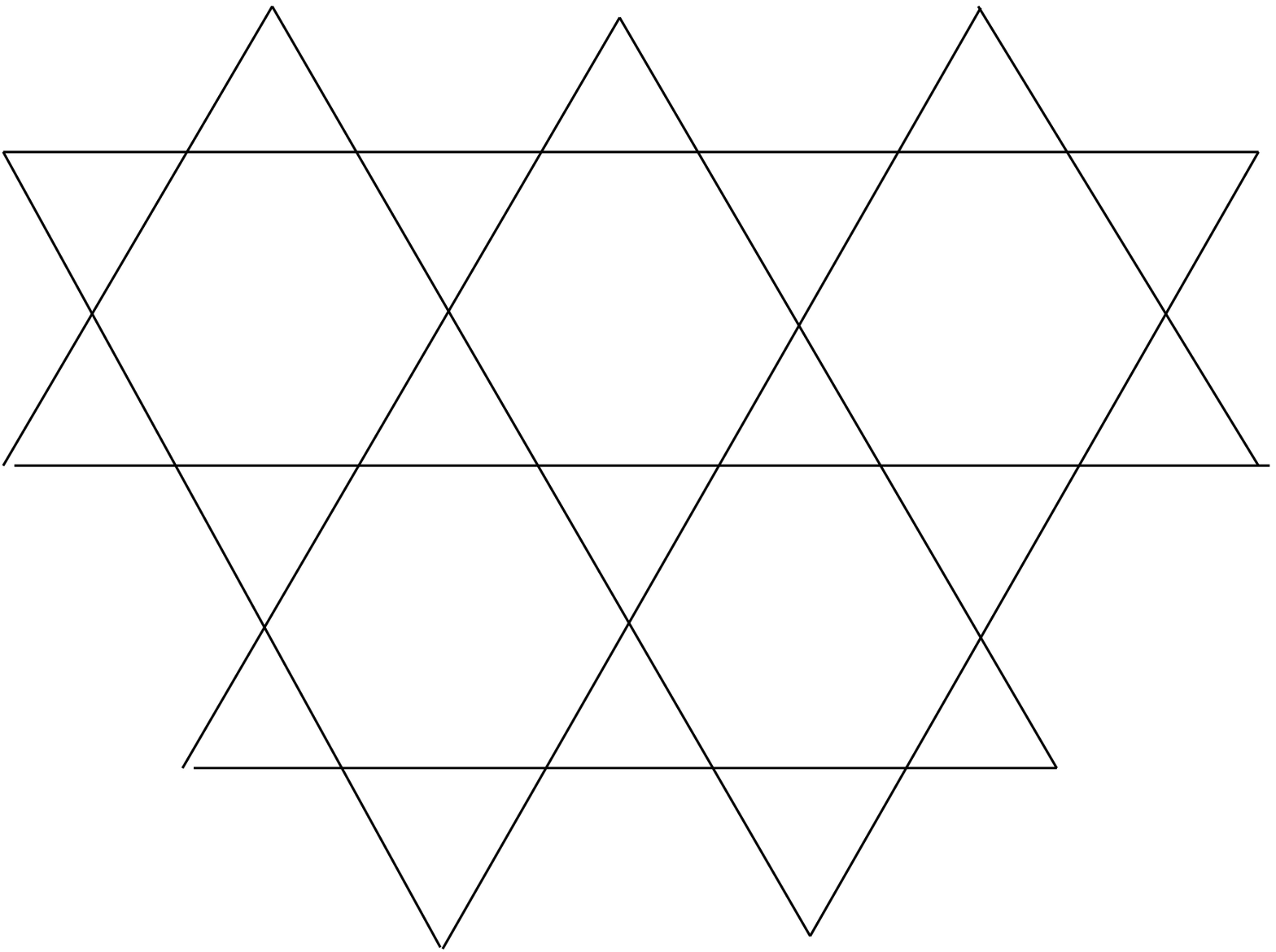}\qquad
\includegraphics*[width=.5\textwidth]{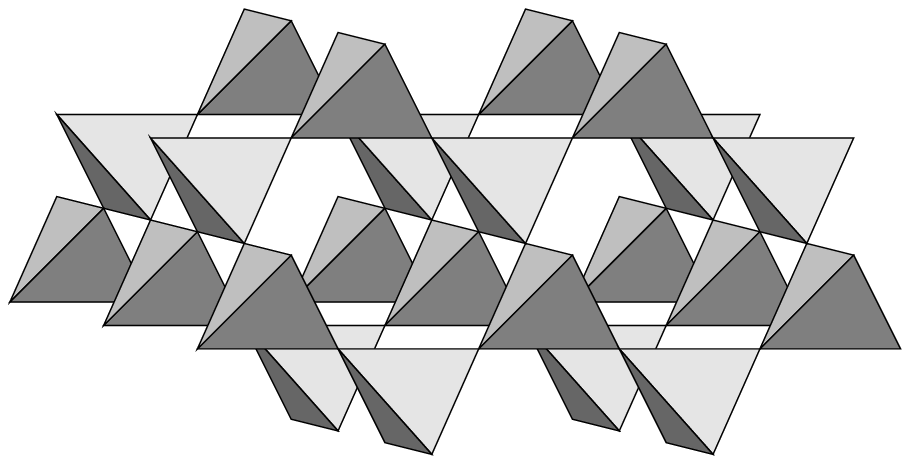}
\caption[]{Left: kagome lattice. Right: pyrochlore lattice}
\label{fig2}       
\end{figure}

The models we are concerned with extend some features present in the
simple system of two corner-sharing triangles to a periodic
lattice. In general we will consider non-bipartite lattices constructed from
corner-sharing arrangements of frustrated clusters, with local magnetic
moments at the vertices  of each cluster and exchange interactions of
equal strength between all moments in each cluster 
(other arrangements are also of interest, but typically show
less dramatic consequences of frustration)
\cite{moessner98,moessner98b}. 
An important
example in two dimensions is the kagome lattice, formed from corner-sharing
triangles; a three-dimensional analogue is the pyrochlore lattice,
built from corner-sharing tetrahedra: see Fig.~\ref{fig2}  for
illustrations of both. 

The Hamiltonian for these models, written in terms of the exchange
energy $J$ and the spin operators ${\bf S}_i$ at sites $i$, has the form
\begin{eqnarray}
{\cal H} = J \sum_{\langle ij \rangle} {\bf S}_i \cdot {\bf S}_j
&\equiv& \frac{J}{2} \sum_{\alpha} \left|{\bf
  L}_{\alpha}    \right|^2 + c\nonumber\\
{\rm where} \qquad
{\bf L}_{\alpha} &=& \sum_{i \in \alpha} {\bf S}_i\;.
\label{H}
\end{eqnarray}
Here the first sum runs over neighbouring pairs $ij$ of sites, while
the second sum is over clusters $\alpha$. To recognise that this
second expression is a correct
rewriting of ${\cal H}$ in
terms of the total spin ${\bf L}_{\alpha}$ of each cluster
$\alpha$, it is necessary only to note that expansion of $|{\bf L}_{\alpha}|^2$ generates
the required off-diagonal terms ${\bf S}_i \cdot {\bf S}_j$, together with
diagonal terms ${\bf S}_i^2$ that contribute to the constant $c$. The apparent
simplicity of this second form is of course deceptive, since the
operators ${\bf L}_{\alpha}$ and ${\bf L}_{\beta}$ associated with two
clusters $\alpha$ and $\beta$ that share a site are not independent.

For future reference it is useful to introduce some terminology. The
frustrated clusters are in general simplices, and their centres
occupy sites of a second lattice, called the simplex lattice. Spins in
our models are located at the mid-points of nearest-neighbour links of
the simplex lattice. For the kagome magnet the simplex
lattice is the honeycomb lattice, and for the pyrochlore magnet it is the diamond
lattice.

While the Hamiltonian of (\ref{H}) provides a useful
basis for understanding the properties of a range of geometrically
frustrated magnetic
materials, various additional physical contributions to a realistic model may
also be important. These include single-ion anisotropy\cite{harris97}, further
neighbour exchange \cite{reimers,reimers91},
dipolar interactions \cite{raju99,palmer00}, Dzyaloshinskii-Moriya
interactions \cite{elhajal05},
magnetoelastic coupling \cite{tchernyshyov02}, site dilution \cite{shender93,moessner99} and exchange
randomness \cite{saunders07}. In
many cases the associated energy scales are small. They set a
temperature scale much smaller than nearest-neighbour exchange, below
which they may induce magnetic order or spin freezing, but they can be
neglected at higher temperatures. We omit all these perturbations
and restrict our discussion to models with only nearest
neighbour exchange.

\section{Some Experimental Facts}
\label{sec:3}

\begin{figure}[t]
\centering
\includegraphics*[width=.4\textwidth]{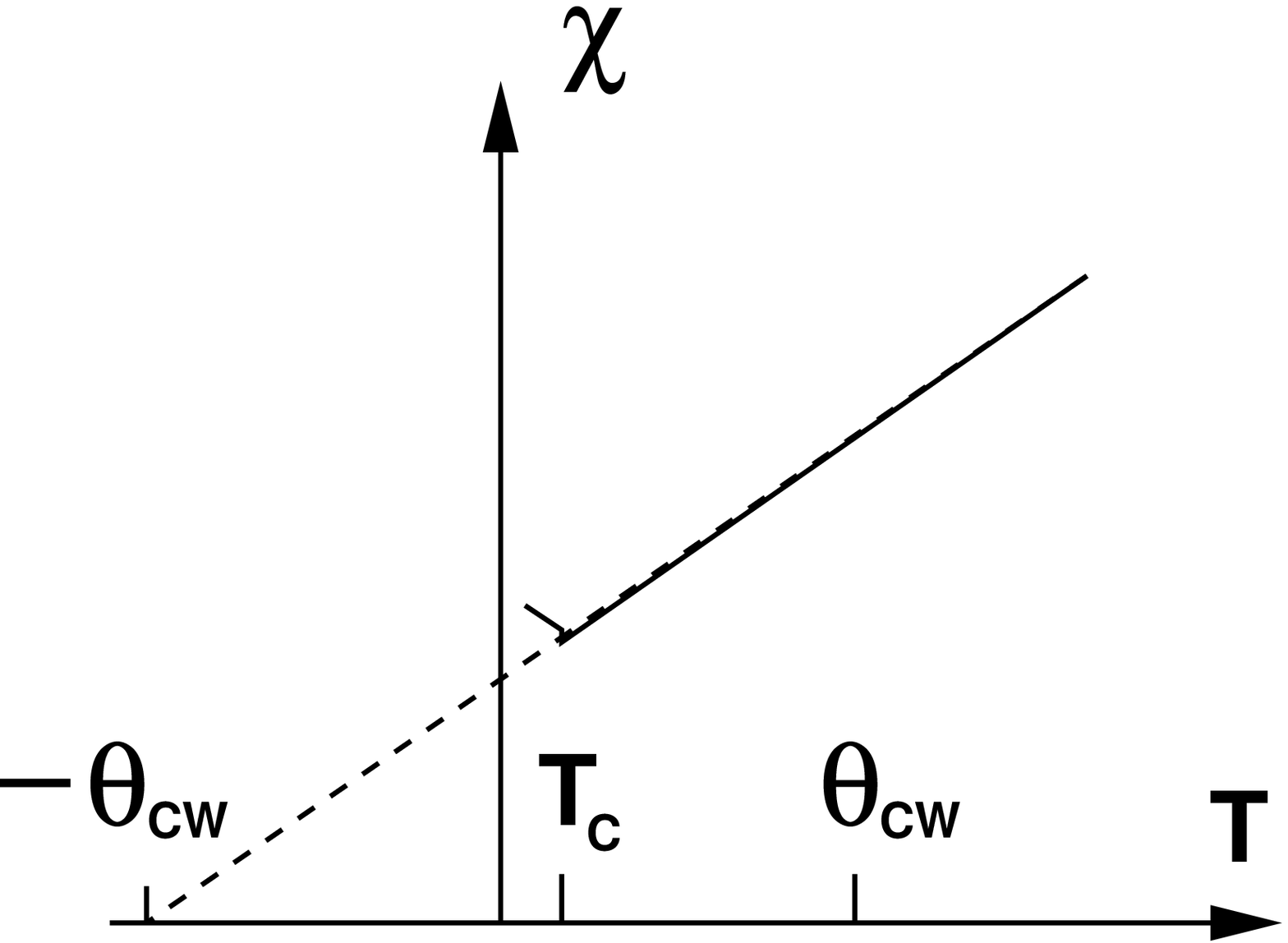} \qquad \qquad
\includegraphics*[width=.4\textwidth]{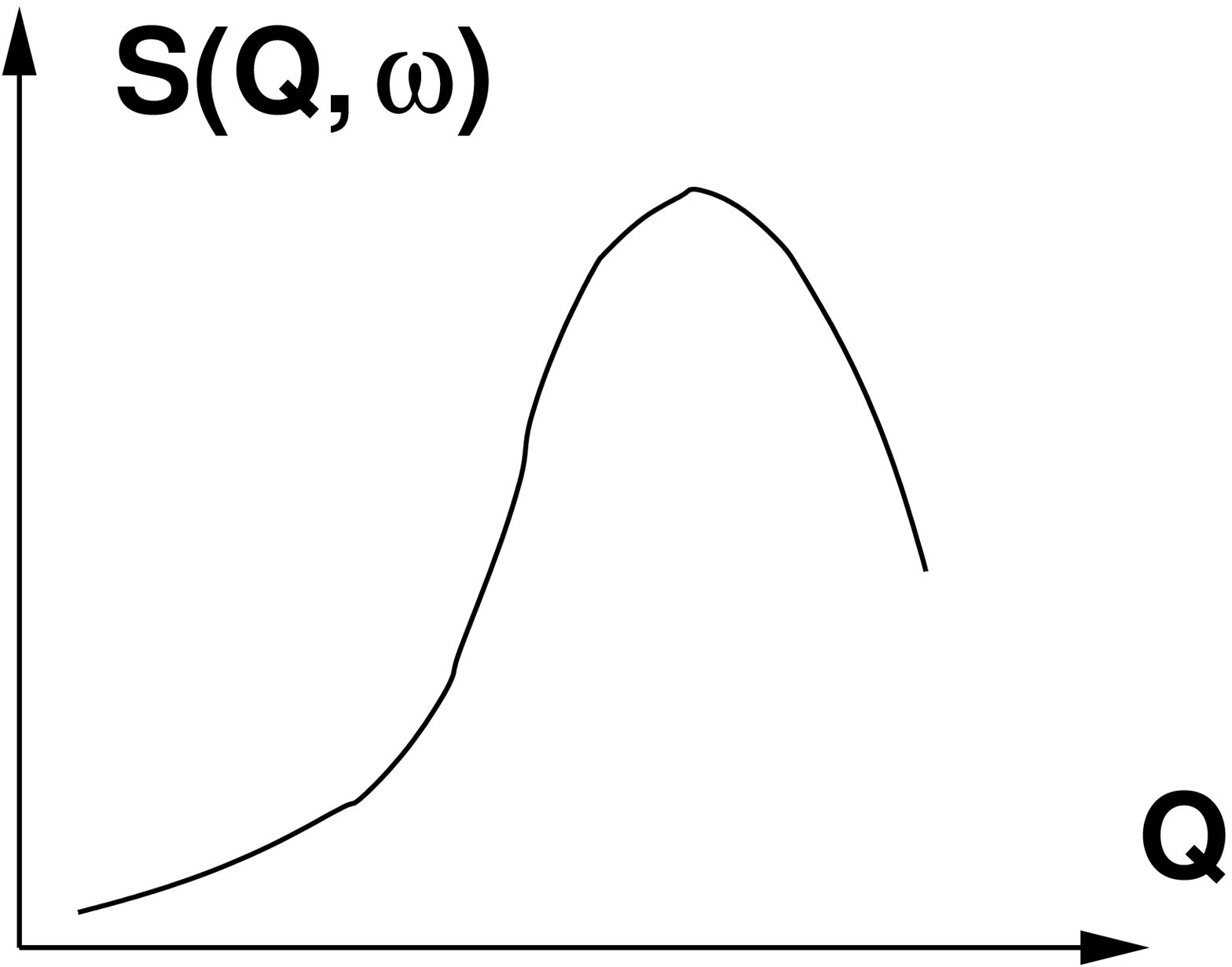}
\caption[]{Characteristic behaviour of a geometrically frustrated
  antiferromagnet. Left: sketch of $\chi^{-1}$ vs $T$.
  Right: sketch of $S(Q,\omega)$ vs $Q$}
\label{fig3}       
\end{figure}

The single most revealing property of a geometrically frustrated
magnet is arguably the dependence on temperature $T$ of its magnetic
susceptibility $\chi$. It is convenient to consider plots of
$\chi^{-1}$ vs $T$, which at high temperature have the linear form
\begin{equation}
\chi^{-1} \propto {T-\Theta_{\rm CW}}\;,
\label{chi}
\end{equation}
where the Curie-Weiss constant $\Theta_{\rm CW}$ characterises the
sign and strength of interactions. In an antiferromagnet
$\Theta_{\rm CW}$ is negative, and for the model of (\ref{H}) 
one has ${\rm k_B}\Theta_{\rm CW} = -zJS^2$. 
Without frustration, magnetic order, signalled by a cusp in $\chi$,
appears below the N\'eel temperature, $T_{\rm N} \sim |\Theta_{\rm
  CW}|$. By contrast, in geometrically frustrated systems nothing
sharp is observed at the temperature scale set by interaction
strength: instead, the paramagnetic phase extends to temperatures
$T\ll \Theta_{\rm CW}$. Ordering or spin freezing may appear at a 
lower temperature $T_{\rm c}$, but a large value for the ratio
$f\equiv|\Theta_{\rm   CW}|/T_{\rm c}$ is a signature of
frustration\cite{ramirez94}. This behaviour is illustrated
schematically in Fig.~\ref{fig3}; references to experimental papers
are given in Table~\ref{tab:1}.

More detailed information on low temperature behaviour is provided by
magnetic neutron scattering (see the chapter by S. T. Bramwell). 
Again, we sketch typical observations in
Fig.~\ref{fig3}, and give references in Table~\ref{tab:1}. The dynamical structure
factor $S(Q,\omega)$ has a broad peak at finite wavevector $Q$,
showing that spin correlations are predominantly short-range and
antiferromagnetic. The width of this peak indicates a correlation
length of order the lattice spacing, while the small value of the elastic
scattering cross-section for $Q\to 0$ shows that correlations suppress
long wavelength fluctuations in magnetisation density.
This form stands in contrast both to that in unfrustrated
antiferromagnets, where N\'eel order leads to magnetic Bragg peaks,
and to that in systems with short-range ferromagnetic correlations, where
the structure factor is peaked at $Q=0$.
Inelastic scattering has a width in frequency $\omega$ that
decreases with decreasing temperature, and in materials that show spin
freezing, scattering weight is transferred from the inelastic to the
elastic response with little change in $Q$-dependence on cooling
through $T_{\rm c}$. 

\begin{table}[h]
\centering
\caption[]{Three geometrically frustrated antiferromagnets}
\renewcommand{\arraystretch}{1.2}
\setlength\tabcolsep{5pt}
\begin{tabular}{@{}lllll@{}}
\hline\noalign{\smallskip}
Material & Structure & $|\Theta_{\rm CW}|$ & $T_{\rm c}$ & References  \\
\hline\noalign{\smallskip}
SrGa$_{3}$Cr$_{9}$O$_{19}$ & pyrochlore slabs & 515 K & 4 K & \cite{ramirez90,broholm90,martinez92,lee96}\\
hydronium iron jarosite & kagome & 700 K & 14 K & \cite{wills98} \\
Y$_2$Mo$_2$O$_7$ & pyrochlore & 200 K & 22 K & \cite{gingras97,gardener99} \\
\hline
\end{tabular}
\label{tab:1}       
\end{table}
Properties of three well-studied geometrically frustrated
antiferromagnets are set out in Table~\ref{tab:1}\footnote{All
  three examples show spin freezing below $T_{\rm c}$}. Two basic
theoretical questions arise. Why is there no magnetic ordering at
$T\sim |\Theta_{\rm CW}|$? And what is the nature of correlations in
the strongly interacting regime $T \ll \Theta_{\rm CW}$?

\section{Classical Ground State Degeneracy}
\label{sec:4}

\begin{figure}[h]
\centering
\includegraphics*[width=.5\textwidth]{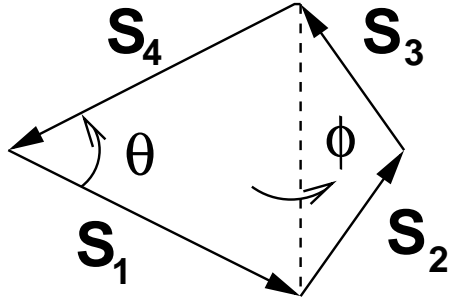}
\caption[]{A ground state configuration for a frustrated cluster of
  four classical Heisenberg spins} 
\label{fig:4}       
\end{figure}

To get insight into the answers to these questions, we start by
considering ground states of models defined by (\ref{H}) in the
classical limit, in which the ${\bf S}_i$ are not operators but three-component vectors of
magnitude $S$. As a first step, it is useful to examine a single
tetrahedral cluster of four spins, with the Hamiltonian
\begin{equation}
{\cal H} = \frac{J}{2} \left|{\bf L} \right|^2 + c \qquad {\rm where}
\qquad {\bf L}={\bf S}_1+ {\bf S}_2+ {\bf S}_3+ {\bf S}_4\;.
\end{equation}
By writing the Hamiltonian in terms of the cluster spin
$\bf L$ we see at once that ground states are those with ${\bf L}={\bf
  0}$. Such an arrangement of four vectors, each having three components,
with resultant zero is shown in Fig.~\ref{fig:4}: these ground
states have two
internal degrees of freedom, indicated in Fig.~\ref{fig:4} by the
angles $\theta$ and $\phi$, in addition to the degeneracies under
global rotations which are expected from the symmetry of $\cal H$.

We should next understand how this accidental ground state degeneracy
extends from a single cluster to a periodic lattice. We can do so
using a counting argument \cite{moessner98,moessner98b}, which
compares $F$, the number of degrees of 
freedom in the system with $K$, the number of constraints that must be
satisfied in ground states. The central point is that if all
constraints are independent, then the number of ground state degrees
of freedom is given by the difference $F-K$. Such an argument was used by
Maxwell in 1864 to discuss the stability of mechanical systems of
jointed rods \cite{maxwell}, and is sometimes referred to as a {\it
  Maxwellian counting argument}. 
For a system of $N_{\rm s}$ classical Heisenberg
spins, $F=2 N_{\rm s}$, since two angles are required to specify the
orientation of each spin. And in a system with the Hamiltonian of
(\ref{H}) consisting of $N_{\rm c}$ clusters, $K = 3 N_{\rm c}$,
since in ground states all three components of ${\bf L}_{\alpha}$ must
be zero for every cluster $\alpha$. Under the assumptions that all
constraints can be satisfied simultaneously, and that they are all
linearly independent, we arrive at an estimate for $D$, the number of
ground-state degrees of freedom: $D=F-K$. Taking the example of the
pyrochlore lattice, we have $N_{\rm s} = 2 N_{\rm c}$ (since four
spins are associated with each tetrahedron, but every spin is shared
between two tetrahedra) and hence $D=N_{\rm c}$, an extensive
quantity. 

This is a striking conclusion: it suggests that there are
local degrees of freedom which can fluctuate independently without the
system leaving its ground state manifold. The argument has two
implications for our understanding of the experimental results
summarised in Sect.~\ref{sec:3}. First, macroscopic degeneracy may prevent
long range order at the temperature scale set by interaction strength,
since there are many low-energy configurations that lack order. Second, since
the magnetisation of each cluster is zero in {\it all} ground states,
the amplitude of long wavelength fluctuations in the magnetisation
density is small at low temperature, and so the dynamical structure
factor $S(Q,\omega)$ is small at low $Q$.

\begin{figure}[hbt]
\centering
\includegraphics*[width=.3\textwidth]{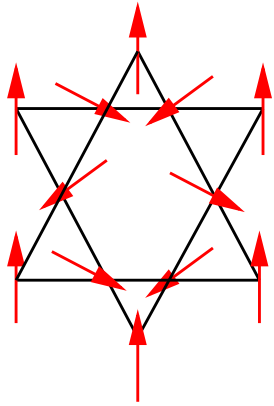}
\caption[]{Illustration of how ground state degrees of freedom arise
  for the Heisenberg model on the kagome lattice: spins on the central
  hexagon may be rotated together through any angle about the axis
  defined by the outer spins, without leaving the ground state. 
}
\label{weathervane-defect}       
\end{figure}
At this point it is worth pausing to consider possible limitations to
the counting argument that has been presented. As noted, it rests on
an assumption that all ground state constraints are linearly
independent. If this is not the case, we underestimate $D$. In our
context, corrections are important if they make an extensive
contribution to $D$. This occurs in the kagome lattice Heisenberg
antiferromagnet: in this case our estimate yields $D=0$ (since, for a
lattice built from corner-sharing triangles, $N_{\rm s} = 3 N_{\rm
  c}/2$), but by explicit construction one finds sets of states with
special spin arrangements
\cite{chalker92,weathervane} for which $D = N_{\rm s}/9$. Such an arrangement is
illustrated in Fig.~\ref{weathervane-defect}. By contrast, for the pyrochlore
Heisenberg antiferromagnet, it is known \cite{moessner98,moessner98b} that
corrections to the estimate for $D$ are at most sub-extensive.

\begin{figure}[h]
\centering
\includegraphics*[width=.7\textwidth]{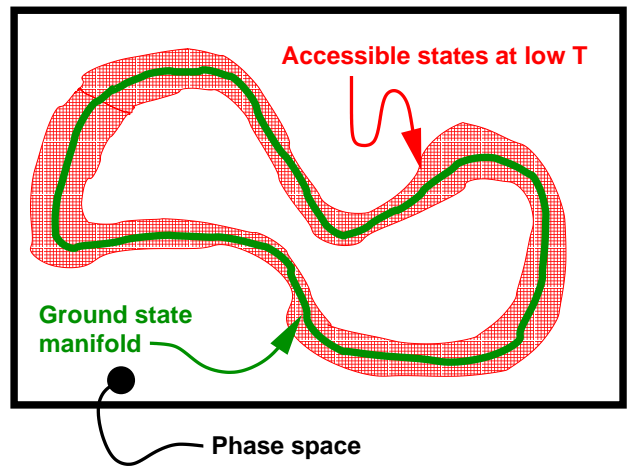}
\caption[]{Schematic view of phase space for a geometrically
  frustrated magnet
}
\label{fig:5}       
\end{figure}

The view of classical geometrically frustrated Heisenberg
antiferromagnets that emerges at this stage is summarised by the
cartoon of phase space given in Fig.~\ref{fig:5}: within
(high-dimensional) phase space for the system as a whole, the ground
states form a manifold with a 
dimension that is much smaller but nevertheless extensive. At 
temperatures small compared to the Curie-Weiss constant (${\rm k_B}T\ll JS^2$),
the system is confined to a region of phase space that 
forms a thin layer around the ground state manifold. Quantum effects
can be neglected provided $JS \ll {\rm k_B}T$, and so a strongly
correlated, classical window, $JS \ll {\rm k_B}T \ll JS^2$, opens for large $S$.

\section{Order by Disorder}
\label{sec:5}

The fact that extensive ground state degeneracy in classical,
geometrically frustrated antiferromagnets is, in the technical sense,
accidental prompts us to ask whether it has robust
consequences in the presence of thermal or quantum fluctuations. 
Specifically, since the degeneracy is not a consequence of symmetry,
one expects the spectrum of fluctuations around each ground state to
be different: the possibility arises that ground states with the
lowest excitation frequencies are selected, because they have the largest
entropy and the smallest zero-point energy. Such an apparently
paradoxical mechanism, by which fluctuations enhance order instead of
suppressing it, is termed `order-by-disorder'
\cite{villain80,shender82}. 

\begin{figure}[h]
\centering
\includegraphics*[width=.4\textwidth]{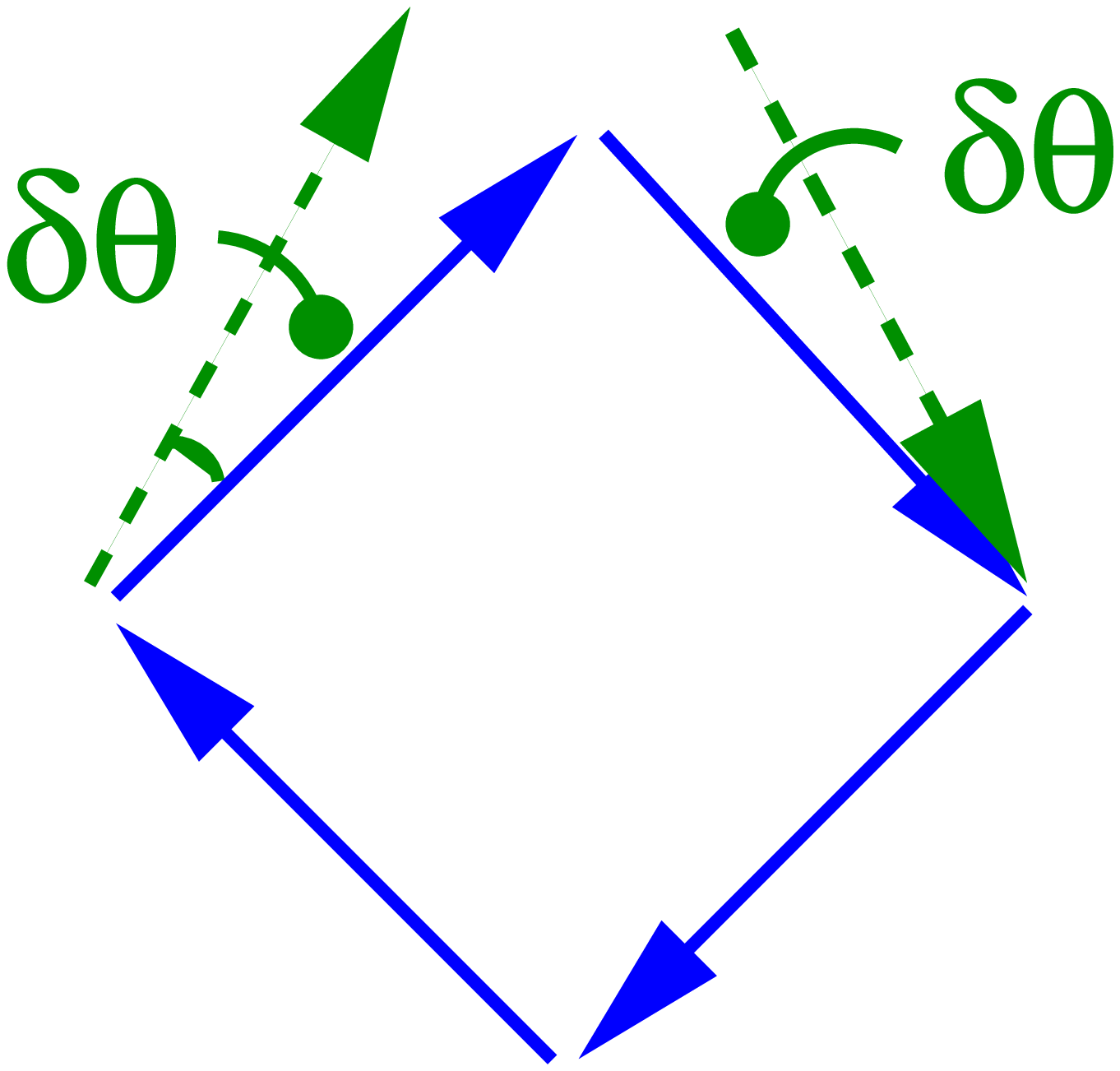}
\includegraphics*[width=.4\textwidth]{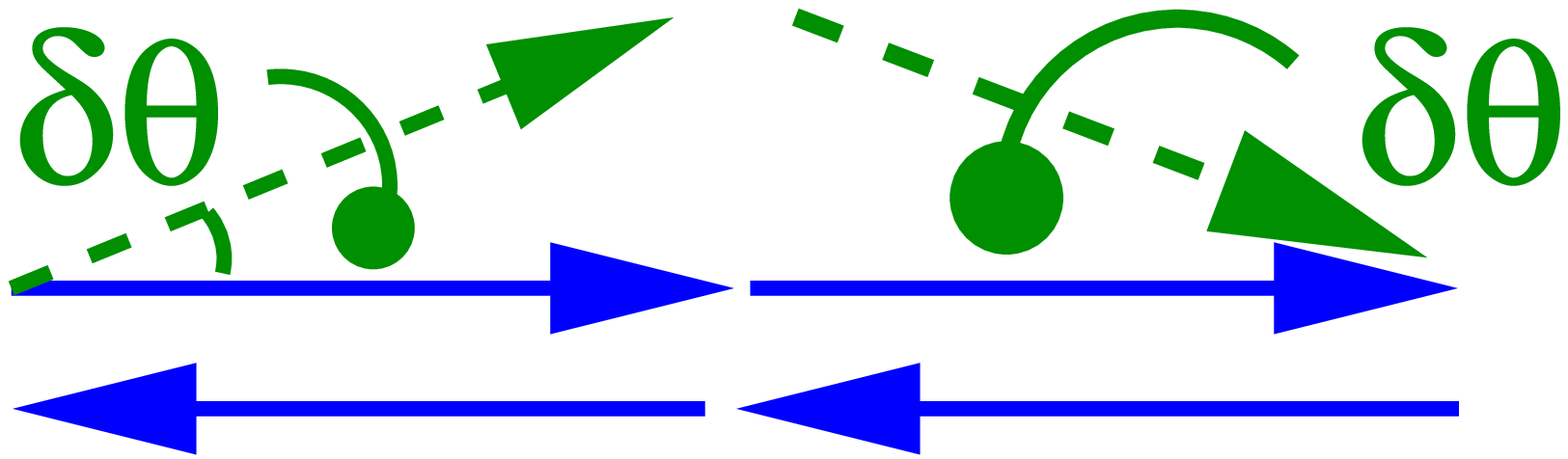}
\caption[]{Fluctuations away from ground state configurations for a
  cluster of four spins. Left: a conventional fluctuation; right: a
  soft mode.
}
\label{fig:6}       
\end{figure}

We will consider first the effects of thermal fluctuations, and begin
by discussing a cluster of four spins. Two ground states with
fluctuations of contrasting types are illustrated in
Fig.~\ref{fig:6}. For the configuration shown on the left, the total
spin of the cluster has a magnitude $|{\bf L}|$ that varies with the
departure $\delta \theta$ from the ground state as $|{\bf L}|\propto
\delta \theta$. Since the excitation energy is proportional to $|{\bf
  L}|^2$, it has a conventional,
quadratic dependence on $\delta \theta$. By contrast, for the
excitation from a collinear ground state shown on the right, $|{\bf
  L}|\propto (\delta \theta)^2$: this mode is therefore soft, with an
energy proportional to $(\delta \theta)^4$. We wish to understand
whether the presence of this soft mode leads almost collinear
configurations to dominate at low temperature.

\begin{figure}[h]
\centering
\includegraphics*[width=.4\textwidth]{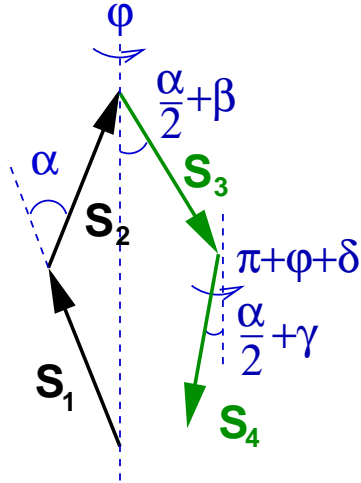}
\caption[]{Coordinate system used for configurations of four spins
}
\label{fig:7}       
\end{figure}

Analysis for a cluster of four spins is simple enough that it can be
followed through in full. To illustrate the range of possible
outcomes, we will consider spins with $n$ components, comparing
behaviour for $n=3$ and $n=2$. We use the coordinate system shown in
Fig.~\ref{fig:7}. Our aim is to evaluate the thermal probability
distribution $P_n(\alpha)$ of the angle $\alpha$ between the pair of
spins ${\bf S}_1$ and ${\bf S}_2$. The distribution
$P_n(\alpha){\rm d}\alpha$ is a product of two factors. One stems from the
measure for ${\bf S}_2$, and is $\sin(\alpha){\rm d}\alpha$ or ${\rm
  d}\alpha$, for $n=3$ or $n=2$ respectively. The other comes from
integrating over orientations of ${\bf S}_3$ and ${\bf S}_4$: it is
\begin{equation}
{\cal Z}_n(\alpha) \propto \int {\rm d}{\bf S}_3 \int {\rm d}{\bf S}_4
\exp\left(-\frac{J}{2T}\left|{\bf S}_3 + {\bf S}_4 - 2S \cos(\alpha/2) \hat{z}\right|^2\right)\;.
\end{equation}
In the low temperature limit, this
can be evaluated by expanding the energy to quadratic order in
deviations from a ground state. For Heisenberg
spins ($n=3)$ the low energy configurations have $|\beta |,\, |\gamma |,\,
|\delta | \ll 1$; for $n=2$, spins are coplanar and two 
coordinates are fixed: $\varphi = \delta = 0$. In a quadratic
approximation the energy is
\begin{equation}
\frac{JS^2}{2}\left\{(\beta - \gamma)^2\cos^2(\alpha/2) + \left[ (\beta + \gamma)^2 +
  \delta^2 \right]
  \sin^2(\alpha/2)\right\}\;, 
\end{equation}
so that (including for $n=3$ a factor of $\sin^2(\alpha/2)$ arising
from ${\rm d}{\bf S}_3 {\rm d}{\bf S}_4$)
\begin{equation}
{\cal Z}_3(\alpha) \propto [\cos(\alpha/2)]^{-1} \qquad
{\rm and} \qquad {\cal Z}_2 \propto [\cos(\alpha/2)\sin(\alpha/2)]^{-1}\;.
\end{equation}
Combining contributions, we have
\begin{equation}
P_3(\alpha) \propto \sin(\alpha/2) \qquad {\rm and} \qquad P_2(\alpha)
\propto \frac{1}{\sin(\alpha)}\;.
\end{equation}

In this way we discover contrasting behaviour for the two cases. With
$n=3$, the system explores all values of $\alpha$ even in the low
temperature limit. But for $n=2$ our unnormalised result for
$Z_2(\alpha)$ has non-integrable divergences at $\alpha = 0$ and
$\alpha = \pi$: in a more detailed treatment, retaining contributions
to the energy quartic in coordinates, these divergences are cut off
on a scale set by temperature, but in the low-temperature limit
$P_2(\alpha)$ approaches a sum of two delta functions, located at $\alpha=0$ and
$\alpha=\pi$.  
Thus order by disorder is absent for $n=3$ but perfect for $n=2$.

\begin{figure}[h]
\centering
\includegraphics*[width=.7\textwidth]{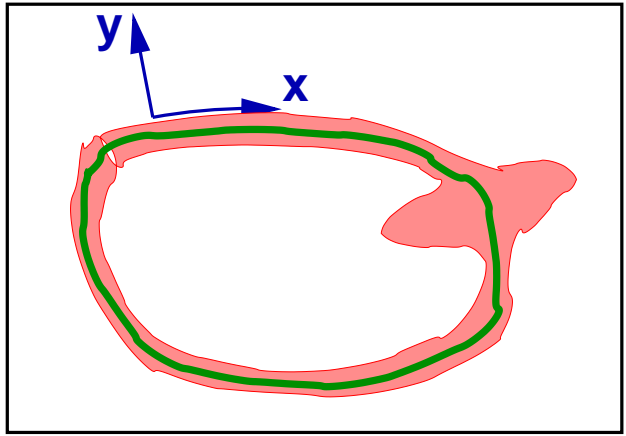}
\caption[]{Schematic view of phase space. The full curve
  represents the ground state manifold. Coordinates $x$ and $y$ are
  respectively parallel and perpendicular to it. 
}
\label{fig:8}       
\end{figure}

Passing from a single cluster to an extended system, consider the
sketch of phase space given in Fig.~\ref{fig:8}. Here, repeating the
convention of Fig.~\ref{fig:5}, the shading indicates the region accessible at low
temperature. One part of this region is concentrated near points on the
ground state manifold at which there are soft modes, as represented by
in Fig.~\ref{fig:8} by a bulge, while another part is distributed in
the neighbourhood of the remainder of the ground state manifold. To
decide whether the system displays order by disorder, we need to
understand which of these two parts dominates. Introducing 
coordinates $\bf x$ and $\bf y$, respectively parallel and
perpendicular to the ground state manifold, in the low temperature
limit we obtain a measure $P({\bf x})$ on the ground state manifold by
integrating over transverse fluctuations
\cite{moessner98,moessner98b}. Characterising these fluctuations by
dynamical frequencies $\omega_l({\bf x})$, we obtain
\begin{equation}
P({\bf x}) \propto \prod_l\left(\frac{{\rm k_B}T}{\hbar \omega_l({\bf x})}\right)\;.
\label{thermal-wt}
\end{equation}
The system has extra soft modes (in addition to those associated with
the ground state coordinates {\bf x})  
at points ${\bf x}_0$ where one or more of the
harmonic frequencies $\omega_l({\bf x}_0)$ vanishes. At these points $P({\bf x}_0)$
is divergent. As for a cluster of four spins, behaviour depends on
whether any such divergence is integrable. If it is, the system
explores the whole of the ground state manifold in the low temperature
limit, but if it is not, then those ground states with soft modes are
selected by thermal fluctuations. Detailed considerations, tested
using Monte Carlo simulations, show for the Heisenberg antiferromagnet
on the kagome lattice that there is coplanar spin order in the low
temperature limit \cite{chalker92}, while on the pyrochlore lattice
there is no order by disorder \cite{moessner98,moessner98b}.

The possibility of ground state selection due to quantum fluctuations
can be discussed using an approach similar in spirit to the one we have taken for
thermal fluctuations, although the outcome has significant differences.
Referring again to Fig.~\ref{fig:8}, one can treat excitations around
a particular point $\bf x$ on the ground state manifold using harmonic
spin wave theory. Excitations involving the coordinates $\bf y$
locally orthogonal to the ground state manifold are conventional modes
with non-zero frequencies $\omega_l({\bf x})$, which have already made an
appearance in (\ref{thermal-wt}). By contrast, fluctuations
involving the coordinates $\bf x$ are, within a harmonic
approximation, zero modes. The zero-point energy of the conventional,
finite-frequency modes provides an effective Hamiltonian for these remaining
degrees of freedom, the classical ground state coordinates. This
Hamiltonian takes the form
\begin{equation}
{\cal H}_{\rm eff}({\bf x}) = \frac{1}{2} \sum_l \hbar \omega_l({\bf
  x})\;.
\label{zero-point-energy}
\end{equation}
The components of $\bf x$ consist of pairs that are, within the
approximations of harmonic spin wave theory (see
(\ref{holstein-primakoff})), 
canonically conjugate.
Treating them as classical commuting variables, the ground state is
the set of points ${\bf x}_{\rm G}$ on which ${\cal H}_{\rm eff}({\bf
  x})$ is minimised. More accurately, the ground state wavefunction
for large $S$ is peaked at ${\bf x}_{\rm G}$, but has zero-point
fluctuations in an effective potential defined by  ${\cal H}_{\rm eff}({\bf
  x})$. 

It is not straightforward to anticipate what features of a classical ground state spin
configuration will minimise ${\cal H}_{\rm eff}({\bf x})$: since all
$\omega_l({\bf x})$ contribute, one could equally imagine focusing on
either the highest frequencies or the lowest ones. In the examples that
have been studied in detail, however, it seems that minima lie at points  ${\bf
  x}_{\rm G}$ where some $\omega_l({\bf x})$ vanish, which we have seen are
also the states favoured by thermal fluctuations. In particular, for
Heisenberg antiferromagnets at large $S$ the selected spin
configurations are coplanar on the kagome lattice \cite{chubukov92} and collinear 
on the pyrochlore lattice \cite{henley06}. In both examples, one third
of the  $\omega_l({\bf x})$ become soft modes at the corresponding points  ${\bf
  x}_{\rm G}$: the coplanar or collinear configurations, respectively.

The principal difference between the ordering effects of thermal and
quantum fluctuations is that in the first case, as we have seen, order
may or may not arise on the limit $J \gg T$, depending on the nature
of the thermal ground state distribution $P({\bf x})$, while in the
second case we always expect order for $S \gg 1$, because by taking $S$
sufficiently large, one can ensure that quantum fluctuations around
the minimum of ${\cal H}_{\rm eff}({\bf x})$ are arbitrarily small.  
Within this framework, the scenario by which one arrives at a spin
liquid on reducing $S$ is clear, at least in principle. For smaller $S$, the
quantum fluctuations are larger and the ground state wavefunction is less well
localised around the minimum of ${\cal H}_{\rm eff}({\bf x})$, while
below a critical value of $S$, the quantum ground state wavefunction
becomes delocalised over the entire classical ground state manifold
and the system loses magnetic order. At large, fixed $S$ long range order
induced by quantum fluctuations is suppressed thermally above a
critical temperature $T_{\rm c} \sim JS$. While the expectation that
spin liquids are favoured at small $S$ is common to our discussions of
both (1.1) and (1.10), one should of course remember that the two equations
embody different physics: harmonic and anharmonic fluctuations,
respectively. 

Efforts to identify experimental examples of order by disorder must
face the problem of establishing that fluctuations, rather than
additional interaction terms in the Hamiltonian, are the cause of what
is observed. For the garnet Ca$_3$Fe$_2$Ge$_3$O$_{12}$, a material
with two interpenetrating magnetic lattices coupled via
zero-point fluctuations, it has been shown that a spinwave gap in the N\'eel
ordered state indeed arises mainly in this way, by independent
determination of the size of single ion anisotropy (the other possible
origin for the gap) \cite{obd-expt1}, and via the characteristic temperature
dependence of the gap \cite{obd-expt2}.

\section{Ground State Correlations}
\label{sec:6}

As we have seen, in some circumstance a model geometrically frustrated
magnet (for example, the classical Heisenberg model on the pyrochlore
lattice) explores its entire ground state manifold at low temperature,
and the experimental evidence from elastic and inelastic neutron
scattering suggests that this is a reasonable picture for the
behaviour of a range of frustrated magnetic materials. We are led to ask in this section whether
there are any important correlations within the ground state manifold.
We will find (for a large class of models: those in which the simplex
lattice is bipartite) that there are indeed
long-range correlations within ground states \cite{moessner98b}, and that these can be
characterised in terms of fluctuations of a Gaussian, divergenceless
field \cite{isakov2004,henley2005}. For this reason, the set of ground
states is said to constitute a Coulomb phase. 

The possibility that spin correlations, averaged over ground states,
have a long-range component, is not self-evident. Indeed, one
might expect the fact that there are a macroscopic number of ground
state degrees of freedom to signal the opposite, since their existence
implies that the set of ground states includes local degrees of
freedom that can fluctuate independently. In turns out, however, that
some ground state correlations are impervious to all local
fluctuations: in this sense they can be said to be topological.

The simplest way to appreciate the existence of long-range
correlations within ground states is to start from (\ref{H}) and
the fact that the total spin ${\bf L}_{\alpha}$ of each frustrated cluster
$\alpha$ vanishes within all ground states.  A consequence of this on the
pyrochlore lattice can be visualised with reference to
Fig.~\ref{fig2}. In particular, consider for any ground state the
total magnetisation ${\bf m}(z)$ of the lowest plane of sites in this
figure. Its value (which may be small -- for example, of order the
square root of the number of sites within the plane, if spins are
randomly orientated within the layer) is perfectly correlated with the
magnetisation of other parallel planes making up the lattice. Indeed,
let ${\bf m}(z+1)$ be the magnetisation of the plane neighbouring the
lowest one. Since the sites in both planes taken together make up a
layer of complete tetrahedra, the overall magnetisation of the layer
is zero, and so ${\bf m}(z+1)=-{\bf m}(z)$. By extension, ${\bf
  m}(z+n)=(-1)^n \times{\bf m}(z)$ for any $n$, a signal of long range
correlations. The correlations give rise to sharp features, termed
{\it pinch points} or {\it bow ties} in the
Fourier transform of the two-point correlation function, averaged over
ground states, as obtained from simulations \cite{zinkin}, large-$n$
calculations \cite{gararin}, and diffuse neutron scattering
measurements \cite{bowties}. These singularities distinguish the diffraction
pattern of the frustrated system from that of a paramagnet, but are
weaker than those of Bragg peaks arising from N\'eel order. 
While their structure can be understood by building on our discussion of
${\bf m}(z)$ \cite{moessner98b}, a more complete
approach uses a long-wavelength description of ground
states.

This continuum description of the Coulomb phase is obtained by mapping
spin configurations onto configurations of vector fields in such a way
that the ground state condition ${\bf L}_{\alpha} = {\bf 0}$,
involving the specifics of the lattice structure, is
translated into the requirement that the vector fields have lattice
divergence zero. This second version of the constraint has the
advantage that it can be implemented in the continuum
\cite{isakov2004,henley2005}. To describe the mapping in 
detail, it is necessary first to discuss some features of the simplex
lattice, introduced in section \ref{sec:2}. We require the simplex
lattice to be bipartite. This is the case, for example, for the diamond
lattice, the simplex lattice associated with the pyrochlore
lattice. (In models without a bipartite simplex lattice, the
correlations discussed in this section are absent \cite{huse2003}.)
For a bipartite simplex lattice, one can adopt an orientation
convention for links, taking them to be directed from the simplices of one
sublattice to those of the other. In this way one can define unit
vectors $\hat{e}_i$ oriented according to the convention and placed at the
mid-points of links of the simplex lattice, which are also the
locations of spins. Considering in the first instance Ising spins, a
spin configuration is represented as a vector field ${\bf B}({\bf r})$ 
on the lattice via
\begin{equation}
{\bf B}({\bf r}_i) = S_i \hat{e}_i\;.
\end{equation}
The condition satisfied in ground states, $\sum_{i \in \alpha} S_i
= 0$, fixes the lattice divergence of ${\bf B}({\bf r})$ to be zero.
More generally, for $n$-component spins ${\bf S}_i$ we require $n$
flavours of vector field ${\bf B}^l({\bf r})$, with $l=1,2 \ldots
n$. These are related to spin components
$S_i^l$ via ${\bf B}^l({\bf r})= S^l_i \hat{e}_i$, so that in ground
states each flavour has divergence zero. 
Note that the fields ${\bf B}^l({\bf r})$ are defined in real space,
and that the global ${\rm O}(n)$ symmetry of the spin Hamiltonian is
implemented as a transformation within the space of flavours, $l$.

Continuum versions of these vector fields ${\bf B}^l({\bf r})$ result
from coarse-graining and the restriction to ground state configurations
is imposed exactly by requiring the continuum fields to be solenoidal.
Each coarse-grained state represents many microscopic
configurations and should have an entropic weight that reflects
this. It is plausible that small continuum field strengths will arise
from many different microscopic configurations, and that large field
strengths will be generated by fewer microscopic states. This suggests
\cite{isakov2004,huse2003} the weight 
\begin{equation}
P[{\bf B}^l({\bf r})] \propto \exp\left(-\frac{\kappa}{2} \int 
{\rm  d}^d{\bf r} \sum_l|{\bf B}^l({\bf r})|^2 \right)\;.
\label{Seff} 
\end{equation}
This theory has a single parameter, the stiffness $\kappa$, whose
value affects the amplitude but not the form of correlations and is
determined microscopically. In $d=3$ dimensions all other
terms consistent with symmetry that might be added to the effective
action are irrelevant in the scaling sense, and so (\ref{Seff}) is
expected to have a universal validity. The resulting correlation
function,
\begin{equation}
\langle B_i^l({\bf 0}) B_j^m({\bf r}) \rangle =
\frac{\delta_{lm}}{4\pi \kappa} 
\left( \frac{3r_i r_j - r^2\delta_{ij}}{r^5}\right)\;,
\label{dipolar} 
\end{equation}
falls off with a fixed, integer power of distance, and has a characteristic,
dipolar angle dependence. 

The fixed, integer power appearing in (\ref{dipolar}) stands in contrast
to behaviour in two other situations in statistical mechanics for which power-law
correlations appear: those of a system undergoing a continuous phase
transition, and of the low-temperature phase in the xy model.
The form of correlations in (\ref{dipolar}) is instead similar to those generated by
Goldstone modes in the ordered phase of a system with a spontaneously
broken continuous symmetry, and an equivalence between that and the
Coulomb phase can be developed by
passing to a dual description of the frustrated magnet
\cite{jaubert}. 

At finite temperature thermal fluctuations out of the ground state
manifold generate a finite correlation length $\xi$ which acts as a
cut-off for the power-law in (\ref{dipolar}). The scale $\xi$
diverges at low temperature, as $\xi \sim T^{-1/2}$ in a Heisenberg model, and
exponentially in an Ising model.

\section{Dynamics}
\label{sec:7}

As we have seen in some detail, geometrically frustrated magnets in
the temperature window $T_{\rm c} \ll T \ll T_{\rm CW}$ are strongly
correlated, yet lack long range order. Their dynamics in this regime
has novel features which we summarise in this section.

\begin{figure}[t]
\centering
\includegraphics*[width=.7\textwidth]{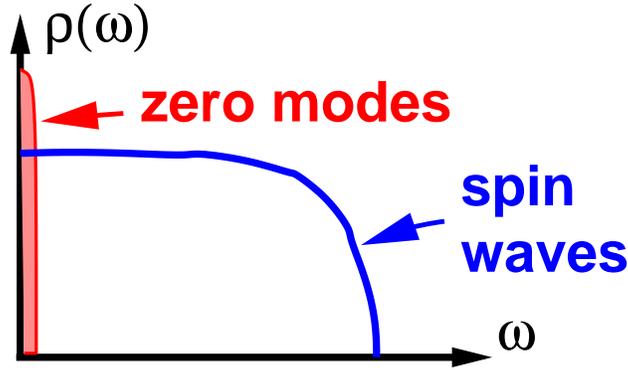}
\caption[]{Density of states in frequency for harmonic excitations in
  a geometrically frustrated antiferromagnet}
\label{fig:9}       
\end{figure}

An obvious first step to understanding low temperature dynamics is
to apply harmonic spinwave theory, starting from one of the ground
states. The results one expects are summarised in terms of the density
of states $\rho(\omega)$ in frequency $\omega$ in
Fig.~\ref{fig:9}. Within the harmonic approximation, 
excitations are of two types. One type, similar to those in
conventional magnets, forms a band of finite-frequency states, with a
maximum frequency $\sim {\cal O}(JS/\hbar)$. The other type (those
associated with the ground state coordinates $\bf x$, in the
discussion of (\ref{zero-point-energy})) are zero modes. For
example, for excitations from a generic ground state of the
Heisenberg antiferromagnet on the pyrochlore lattice, one quarter are
zero modes. 

There is a clear interest in understanding in more detail the
nature of these zero modes. In the most cases, however there is a
obstacle to a simple, analytical treatment, which stems from the fact
that spin configurations in representative classical ground states do
not have long range order. This means that, even though the lattice
itself is periodic, the equations of motion cannot be diagonalised by
Fourier transform, and results of the kind sketched in
Fig.~\ref{fig:9} can be obtained only numerically.

To circumvent this difficulty and illustrate in a simple fashion how
a dispersionless band of modes can arise, it is interesting to consider
a geometrically frustrated Heisenberg antiferromagnet in the presence
of a magnetic field $h$ strong enough that the ground state is fully
polarised. Using the standard Holstein-Primakoff transformation to
write operators for spin components $S_i^l$ in terms of boson creation
and annihilation operators $a^{\dagger}_i$ and
$a^{\phantom{\dagger}}_i$,
with
\begin{eqnarray}
S_i^z &=& S - a^{\dagger}_i a^{\phantom{\dagger}}_i \nonumber \\
S_i^+ &=& (2S)^{1/2} a^{\phantom{\dagger}}_i + \ldots \nonumber \\
S_i^- &=& (2S)^{1/2} a^{\dagger}_i + \ldots \;,
\label{holstein-primakoff}
\end{eqnarray}
we have
\begin{equation}
{\cal H} = J\sum_{ij} {\bf S}_i \cdot {\bf S}_j - h \sum_i S_i^z = JS
\sum_{ij} \left[ a^{\dagger}_i a^{\phantom{\dagger}}_j +
  a^{\dagger}_j a^{\phantom{\dagger}}_i\right] - \mu \sum_i
a^{\dagger}_i a^{\phantom{\dagger}}_i + {\cal O}(S^0)\;.
\label{boson}
\end{equation}
The right-hand form of (\ref{boson}) is a tight-binding model for bosons moving on
the lattice with a nearest-neighbour hopping amplitude $JS$ and a chemical potential
$\mu\equiv zJS-h$ that is linear in the magnetic field $h$. It is a characteristic
of the lattices we are concerned with that such a tight-binding model
has a dispersionless band with eigenvalue $-2JS$, which lies at the
bottom of the spectrum for $J>0$. 

\begin{figure}[h]
\centering
\includegraphics*[width=.5\textwidth]{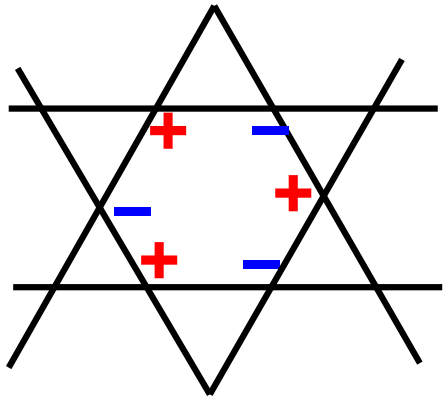}
\caption[]{A magnon mode from the dispersionless band 
}
\label{fig:10}       
\end{figure}

Eigenvectors of the tight-binding Hamiltonian from the dispersionless band are
straightforward to picture. One for the kagome lattice is represented
in Fig.~\ref{fig:10}. Here, eigenvector amplitudes are zero at all sites
except for those around one hexagon of the lattice, on which they have
equal magnitude and alternating signs. The state is an eigenvector
because there is destructive interference between hopping processes that move
the boson off the occupied hexagon. It belongs to a dispersionless band
since it is degenerate with many other, equivalent states, based on
the other hexagons of the lattice. The condition for an arbitrary
vector, with site amplitudes $\psi_i$, to be a linear superposition of
such states is that $\sum_{i \in \alpha} \psi_i = 0$ for each triangle
$\alpha$. Both the extension of this condition to other lattices
constructed from corner-sharing simplices and its parallel with the
ground state condition in spin models, ${\bf L}_{\alpha} = {\bf 0}$
for all $\alpha$, are obvious. There is a gap to excitations for large
$h$, when $\mu$ is large and negative. The gap falls to zero at the critical
field strength $h_{\rm c} = JS(z+2)$ at which $\mu$ crosses the energy
of the dispersionless magnon band. For $h<h_{\rm c}$ these states are populated, and the
magnetisation deviates from its saturated value. In this field range
we recover the classical ground state degeneracy of the zero-field
problem.

We now turn to a discussion of 
dynamics beyond the harmonic approximation. As a starting point we
should consider the full equation of motion
\begin{equation}
\hbar \frac{{\rm d} {\bf S}_i}{{\rm d}t} = {\bf S}_i \times {\bf H}_i
= \frac{J}{2}\left\{\left[ {\bf L}_{\alpha} + {\bf L}_{\beta}
\right]\times  {\bf S}_i - {\bf S}_i \times \left[ {\bf L}_{\alpha} + {\bf L}_{\beta}
\right] \right\}\;.
\label{eqn-of-motion}
\end{equation}
Here ${\bf H}_i$ is the exchange field acting at site $i$. It is
parallel to ${\bf S}_i$ in a classical ground state, but in excited
states has transverse components. These can
be expressed as shown, in terms of the total magnetisations ${\bf L}_{\alpha}$
and ${\bf L}_{\beta}$ of the clusters $\alpha$ and $\beta$ that share
the site $i$. 
For large $S$ we can treat this equation
of motion classically. Within the harmonic approximation, obtained by
linearising the right-hand side, the exchange field ${\bf H}_i$ is a
superposition of contributions from finite frequency modes, which
maintain phase coherence indefinitely, and therefore average to zero over long
times. Anharmonic interactions have two consequences, which are
distinct but turn out to be closely linked \cite{moessner98,moessner98b}.  
One is to generate a lifetime for the finite frequency modes. The
other is to introduce coupling between the ground state coordinates
$\bf x$ and the coordinates $\bf y$ orthogonal to the ground state
manifold. Over long timescales this coupling drives the system around
the ground state manifold. Within the framework of
(\ref{eqn-of-motion}), this long-time component to the dynamics
arises because, once spinwaves have a finite lifetime, the
exchange field is no longer a superposition of exactly harmonic
contributions. Instead, on timescales longer than the lifetime, it is
better thought of as a stochastic quantity. In turn, the long-time
motion of the system around the ground state manifold is itself a
source of dephasing for finite frequency excitations. Specifically,
since the harmonic Hamiltonian is time-dependent, an adiabatic
approximation is not exact, and modes are mixed at long times.
There is a separation of timescales, since typical spinwave periods
are fixed, while spinwave lifetimes diverge as $T^{-1/2}$ and the
timescale for motion between groundstates diverges faster, as $T^{-1}$
\cite{moessner98b}.

These ideas suggest a much simpler approach to calculating the spin
auto-correlation function, in which we treat the exchange field as a
stochastic quantity, using the equation of motion 
\begin{equation}
\frac{{\rm d}{\bf S}(t)}{{\rm d}t} = {\bf S}(t) \times {\bf H}(t)
\end{equation}
with the correlation function 
\begin{equation}
\langle H^l(t) H^m(t^{\prime}) \rangle = \Gamma \delta_{lm}
\delta(t-t^{\prime})
\label{langevin} 
\end{equation}
for components $H^l(t)$ of ${\bf H}(t)$.
The noise intensity $\Gamma$ can be estimated using equipartition. 
We have
\begin{equation}
\Gamma = \int_{-\infty}^{\infty} \langle {\bf H}_i(0) \cdot
{\bf H}_i(t)\rangle \,{\rm d}t\,
 \sim  \int_{-\infty}^{\infty} \langle {\bf L}_{\alpha}(0)
\cdot {\bf L}_{\alpha}(t)\rangle\, {\rm d}t\;.
\end{equation}
In addition, ${\bf L}_{\alpha}(t)$ is a superposition of contributions with
amplitudes $A_{\omega}$ from thermally excited spinwaves: 
\begin{equation}
{\bf L}_{\alpha}(t) = \sum_l A_{\omega_l} {\rm e}^{-{\rm i} \omega_l t}\;.
\end{equation}
Combining these assumptions, we find \cite{moessner98b}
\begin{equation}
\Gamma \sim \left. \langle |A_\omega|^2 \rangle \rho(\omega)
\right|_{\omega \to 0} \sim \frac{k_{\rm B} T}{JS}\;.
\end{equation}
The Langevin equation (\ref{langevin}) itself is straightforward
to solve, and yields
\begin{equation}
\langle {\bf S}(0) \cdot {\bf S}(t)\rangle = S(S+1) \exp(-ck_{\rm
  B}Tt/\hbar S)\;,
\label{autocorrelation}
\end{equation}
where $c \sim {\cal O}(1)$ is an undetermined numerical constant.
This result is notable for the fact that temperature alone sets the
time scale: $J$ drops out of the long-time dynamics in the low
temperature regime. In this sense, behaviour matches that expected at
a quantum critical point, although the underlying physics is quite different.

The predictions of (\ref{autocorrelation}) have been tested both
in simulations and in experiment.
Molecular dynamics simulations proceed by 
direct integration of the equations of motion,
(\ref{eqn-of-motion}), with an initial configuration drawn from a
thermal distribution and generated via Monte Carlo simulation.
Results for the classical pyrochlore Heisenberg antiferromagnet
\cite{moessner98,moessner98b}
in the temperature range ${\rm k_B}T \ll JS^2$ reproduce both the functional form
of (\ref{autocorrelation}) for the time dependence of the
autocorrelation function and the scaling of relaxation rate with temperature. 
In experiment, inelastic neutron scattering offers direct access to
spin dynamics. Early measurements of the energy width of quasielastic
scattering in CsNiFe$_6$ \cite{harrisneutron2} were fitted to a Lorentzian, the
transform of the time-dependence given in
(\ref{autocorrelation}), yielding a 
relaxation rate for $T< |\Theta_{\rm CW}|$ that is strongly
temperature dependent,
although without specific evidence for the (subsequently proposed)
linear variation with $T$. More detailed data for SCGO \cite{aeppli} 
confirms a relaxation rate of order $k_{\rm B}T/\hbar$, and very
recent measurements on Y$_2$Ru$_2$O$_7$  \cite{broholm2007} display
rather clearly a relaxation rate proportional to temperature. For the
future, it would be interesting to have both theoretical and
experimental information not simply on the autocorrelation function,
but on dynamics as a function of wavevector.

\section{Final Remarks}
\label{sec:8}

In conclusion, we have seen how geometrical frustration in classical
magnets can lead to macroscopic ground state degeneracy and the
suppression of long range order. Low temperature states in model
systems, although disordered, are very different from those of a
non-interacting paramagnet: correlations are power-law in space, and
decay in time at a rate set by temperature alone. Many experimental
systems display these features within the temperature window $T_{\rm c}
< T < |\Theta_{\rm CW}|$ where behaviour is dominated by nearest
neighbour exchange. Behaviour in this regime is well summed up in the term coined by
Jacques Villain \cite{villain79}, a pioneer in the field: {\it Cooperative Paramagnetism}.

\subsubsection{Acknowledgements}

I am very grateful to my collaborators in the research on
geometrically frustrated antiferromagnets that I have been involved
with, especially E. F. Shender, P. C. W. Holdsworth and R. Moessner,
and also J. F. G. Eastmond, S. E. Palmer, P. Hogan, M. Y. Veillette,
R. Coldea, T. E. Saunders, M. J. Bhaseen and T. S. Pickles. In addition, I have benefitted
from discussions with many other colleagues. I thank EPSRC for
supporting this work.

%
\input{chalker-references.tex}



\printindex
\end{document}

%% file: chalker-references.tex
%
%

%
%